\documentclass[
    11pt,
    letterpaper,
    reprint,
    notitlepage,
    superscriptaddress,
    aps,
    prl
]{revtex4-2}

\usepackage[dvipsnames]{xcolor}
\usepackage{graphicx}
\usepackage{bm}
\usepackage{physics}
\usepackage{mathtools}

\usepackage{amsmath,amssymb,amsfonts} 
\usepackage[makeroom]{cancel}
\usepackage{hyperref}
\hypersetup{colorlinks=true,citecolor=blue,linkcolor=blue,urlcolor=blue}

\newcommand*{\Omq}{\ensuremath{\Omega_{\rm{q}}}}
\newcommand*{\ns}{\ensuremath{\hat n_{\rm{S}}}}
\newcommand*{\nf}{\ensuremath{\hat n_{\rm{F}}}}
\newcommand*{\gammam}{\ensuremath{\gamma_{\rm{m}}}}
\newcommand*{\omegam}{\ensuremath{\omega_{\rm{m}}}}
\newcommand*{\Omegaf}{\ensuremath{\Omega_{\rm{F}}}}
\newcommand*{\Omegas}{\ensuremath{\Omega_{\rm{S}}}}

\begin{document}

\title{
Universality of stationary entanglement in an optomechanical system driven by non-Markovian noise and squeezed light}

\author{Su Direkci}
\email{sdirekci@caltech.edu}
\affiliation{The Division of Physics, Mathematics and Astronomy,
California Institute of Technology, CA 91125, USA} 
\author{Klemens Winkler}
\affiliation{Vienna Center for Quantum Science and Technology (VCQ), Faculty of Physics, University of Vienna, A-1090 Vienna, Austria}
\author{Corentin Gut}
\thanks{Present address: KEEQuand GmbH, Gebhardtstr. 28, 90762 Fürth, Germany.}
\affiliation{Vienna Center for Quantum Science and Technology (VCQ), Faculty of Physics, University of Vienna, A-1090 Vienna, Austria}
\author{Markus Aspelmeyer}
\affiliation{Vienna Center for Quantum Science and Technology (VCQ), Faculty of Physics, University of Vienna, A-1090 Vienna, Austria}
\affiliation{Institute for Quantum Optics and Quantum Information (IQOQI) Vienna, Austrian Academy of Sciences, Boltzmanngasse 3, 1090 Vienna, Austria} 
\author{Yanbei Chen}
\affiliation{The Division of Physics, Mathematics and Astronomy,
California Institute of Technology, CA 91125, USA}

\date{\today}

\begin{abstract}
Optomechanical systems subjected to environmental noise give rise to rich physical phenomena. We investigate entanglement between a mechanical oscillator and the reflected coherent optical field in a general, not necessarily Markovian environment. For the input optical field, we consider stationary Gaussian states and frequency-dependent squeezing.
We demonstrate that for a coherent laser drive, either unsqueezed or squeezed in a frequency-independent manner, optomechanical entanglement is destroyed after a threshold that depends only on the environmental noises---independent of the coherent coupling between the oscillator and the optical field, or the squeeze factor. In this way, we have found a universal entangling-disentangling transition.
We also show that for a configuration in which the oscillator and the reflected field are separable, entanglement cannot be generated by incorporating frequency-dependent squeezing in the optical field.
\end{abstract}

\maketitle

Quantum theory predicts that an object becomes correlated with its measurement apparatus during a measurement process, leading to their mutual entanglement---regardless of the details of the interaction and the size of the joint object-apparatus system~\cite{Neumann_1955}. 
%
However, interactions with the environment can cause decoherence within the object-apparatus system, even destroying their entanglement, as often happens 
%
in the  macroscopic regime~\cite{zurek_pointer, Joos_Zeh_1985, schlosshauer_book, SCHLOSSHAUER2019_decoherence}. 

Optomechanical systems, formed by an optical field interacting with a mechanical object, are promising platforms for exploring quantum phenomena in the macroscopic realm 
\cite{GENES200933, TANG_perspective_2022}. 
Due to the high quality factor of the mechanical oscillator, these devices ensure strong isolation from the environment and maintain quantum coherence \cite{Aspelmeyer_cavity_optomechanics}. As a key feature of quantum coherence, optomechanical entanglement between 
light and 
mechanical motion has been studied \cite{hofer_2011} and experimentally observed in a pulsed regime \cite{palomaki_2013}. Stationary entanglement, which arises when the object is measured continuously by a continuum of light modes coupled to the mechanical object, has been studied theoretically and proposed for experimental demonstration~\cite{GENES200933, miao_universal_2010, miao_probing_macroscopic, gut_stationary_2020, direkci_2024, PhysRevLett.99.250401,PhysRevLett.98.030405}.

This Letter and its companion Article \footnote{See the companion Article titled ``Characterizing Stationary Optomechanical Entanglement in the Presence of Non-Markovian Noise''.} answer a general question about optomechanical entanglement: what determines the presence of stationary light-mass entanglement? Is it (i) the quality of isolation from the environment, (ii) the optomechanical interaction strength, and/or (iii) the state of the incoming light field? These factors are intimately tied to the sensitivity of the optomechanical system to weak forces acting on the mass: (i) determines the {\it environmental noise} of the device, while (ii) and (iii) determine its \emph{quantum noise}, i.e. the unavoidable, fundamental measurement noise of the light measuring the position of the mass. In fact, our work has been directly motivated by the achievement of quantum noise below the \emph{standard quantum limit} (SQL) \cite{Danilishin_qm_theory,chen_macroscopic_2013} by the Laser Interferometer Gravitational-Wave Observatory (LIGO) \cite{advanced_ligo}, via the injection of optical squeezed states ~\cite{yu2020quantum,lee_ligo_freq_squeezing}.  
Our short answer is that factor (i) plays a decisive role, while (ii) and (iii), as important as they are for sensing weak forces, cannot be used to bring optomechanical entanglement into existence. However, once (i) allows for entanglement, (ii) and (iii) can be used to tune the level of entanglement.

We consider a single oscillator monitored by a continuous beam of light. The joint system suffers from two types of stationary, Gaussian environmental noise sources that are non-Markovian in general: a \emph{force noise} acting on the center-of-mass of the mechanical mode, and a \emph{sensing noise} acting on the reflected light (it characterizes the difference between the center-of-mass position and the position that the light senses). The input light contains a single carrier field, with vacuum or stationary, squeezed fluctuations.

\begin{figure}
    \centering
\includegraphics[width=\linewidth]{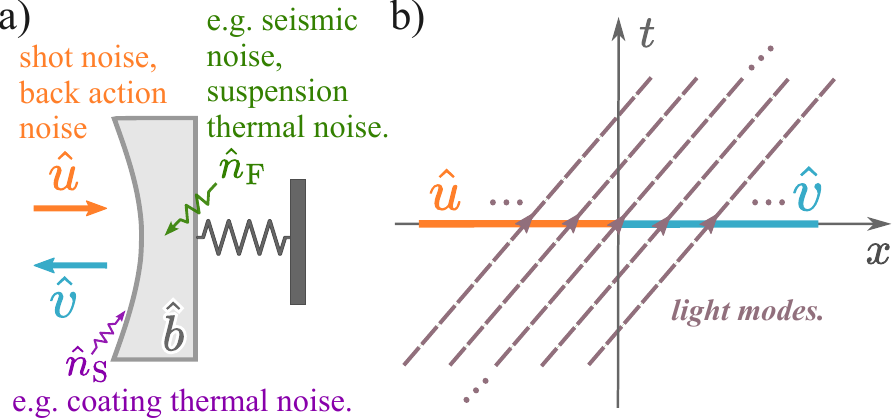}
    \caption{Schematic of the optomechanical system. a) The input light field, $\hat u$, interacts with the mechanical mode, $\hat b$, and is reflected. We refer to the reflected light as the output light field, $\hat v$. The system is also subjected to environmental noise sources $\hat{n}_{\textrm{F}}$ and $\hat{n}_{\textrm{S}}$. b) Space-time diagram of the system. We choose the convention that the light field is traveling from $x \rightarrow -\infty$ to $x = 0$ before it interacts with the oscillator. Afterwards, it travels to $x \rightarrow \infty$. The equivalence between temporal and spatial modes can be seen from the diagram: the input field can be thought of as the spatial modes for $x < 0$ at $t=0$ (shown in orange), or the temporal modes for $t>0$ at $x=0$. Similarly, the output field comprises the spatial modes for $x > 0$ at $t=0$ (shown in blue), or the temporal modes for $t<0$ at $x=0$.}
    \label{fig:light_curves}
\end{figure}

In particular, when the incoming light field is the vacuum or a frequency-independently squeezed state, we find that stationary optomechanical entanglement exists for any level of force noise, in the absence of sensing noise.
When a sensing noise is present, 
the transition from an entangled state to a separable one is independent of the strength of the optomechanical interaction, or the squeezing in the light field: We call this effect the \emph{universality of optomechanical entanglement}, which has been observed  numerically for specific cases involving both Markovian and non-Markovian noises in \cite{direkci_2024}, and proved rigorously in the companion Article \footnotemark[1]. Note that here, we consider sensing noises that are measured simultaneously with the mechanical position. The universality of optomechanical entanglement vanishes in the presence of other types sensing noises affecting the light field, such as passive losses or dark noise, which would also increase the difficulty in observing optomechanical entanglement in a practical setting \cite{optomech_entanglement_photon_counting}. For an example calculation, see Sec. III.C in \footnotemark[1].

If the incoming light field is frequency-dependently squeezed, the entangling-disentangling transition is no longer universal, and depends on the interaction strength. However, we prove that, when environmental noise levels are kept fixed, frequency-dependent squeezing always makes optomechanical entanglement harder to achieve compared to the vacuum case. Furthermore, if the system is separable for the vacuum input, it is bound to be separable for any frequency-dependent squeezing. 

\emph{System Dynamics---} We model our mechanical mode as a single harmonic oscillator described by its dimensionless  position and momentum quadratures,  $\hat b_1$ and $\hat b_2$, with $[\hat b_1, \hat b_2]= 2i$ \footnote{The factor of $2$ is a particular normalization that simplifies the derivations, and it is found in several works with continuous variables, see \cite{cv_qi_adesso} for example}. We model the optical field traveling towards the mechanical oscillator (input field) with Hermitian amplitude and phase quadratures $\hat u_1(t)$ and $\hat u_2(t)$, respectively. The continuous index $t$ labels the infinitely many temporal modes of the incoming light field: they correspond to ``rays of light" that interact with the mechanical mode at time $t$ (see Fig. \ref{fig:light_curves}). Similarly, the light field reflected from the oscillator (output field) at time $t$ is described by Hermitian amplitude and phase quadratures $\hat v_1(t)$ and $\hat v_2(t)$, respectively. These quadratures obey equal-time bosonic commutation relations $[\hat q_j(t), \hat q_k(t')] = 2 i \delta_{jk} \delta(t-t')$,
with $q = u, v$, $j,k=1,2$. Our system is governed by the following phenomenological quantum Langevin equations \cite{Muller-Ebhardt_2008}:
\begin{subequations}
 \label{eqn:eqns_time_domain}
\begin{align}
\hat{v}_1(t) &= \hat{u}_1(t) \,, \\
\hat{v}_2(t) &= \hat{u}_2(t) + {\Omq}\omegam^{-1/2} \big[\hat{b}_1(t)+\ns(t)\big] \,, \label{eqn:v2(t)} \\
 \dot{\hat{b}}_2(t) &= -\gammam \hat{b}_2(t) -\omegam \hat{b}_1(t) +{\Omq}\omegam^{-1/2}  \hat{u}_1(t) +\nf(t), \label{eqn:b2dot} \\
\dot{\hat{b}}_1(t) &= \omegam \hat{b}_2(t) \,.\label{eqn:b1dot}
\end{align}
\end{subequations}
These linearized equations are valid when the input light is highly excited (e.g. for a high intensity laser $\ket{\alpha}$, with $\alpha \gg 1$). We also choose a displaced frame rotating with the laser frequency, such that the first moments of all of the operators vanish at all times \cite{Aspelmeyer_cavity_optomechanics, hofer_thesis, adesso_entanglement_2007}. The coherent optomechanical interaction has a strength of $\Omega_{\rm{q}}$ in units of frequency \footnote{In terms of system parameters, $\Omq = \sqrt{\frac{I_{\rm{probe}} \omega_{\rm{probe}} }{M c^2}}$, where $I_{\rm{probe}}$ and $\omega_{\rm{probe}}$ are the intensity and the frequency of the light probing the mechanics, $c$ is the speed of light in vacuum, and $M$ is the mass of the oscillator.}, which we refer to as the \emph{interaction strength}, or the coherent optomechanical coupling. The mechanical oscillator has a resonance frequency of $\omegam$, and a viscous damping rate of $\gamma_{\rm{m}}$. Then, an external force noise $\hat n_{\rm{F}}$ acts on the oscillator. Instead of considering a thermal bath, where the spectrum of $\hat n_F$ is given by the
fluctuation-dissipation theorem \cite{Kubo_1966, callen_1951},
we consider phenomenologically a general, non-Markovian, and Gaussian stationary noise with a power spectral density in the form of
\begin{equation}
\label{eqn:noise_rational}
    S_{n_{\rm{F}}}(\Omega) \coloneqq \int_{-\infty}^\infty dt \expval{\hat n_{\rm{F}}(t) \hat n_{\rm{F}}(0)} e^{i \Omega t} = {P(\Omega)}/{Q(\Omega)}
\end{equation}
where $P$ and $Q$ are arbitrary polynomials in $\Omega$ such that $\text{deg}(Q) \geq \text{deg}(P)$ due to stability requirements. 

The measurement of the oscillator position $\hat b_1$ is enabled by the linear, momentum-exchange optomechanical interaction $\hat u_1 \hat b_1$~\cite{zurek_pointer, schlosshauer_book, Wiseman_Milburn_control, hofer_thesis}. 
This leads to the asymmetrical Langevin equations where only the oscillator's momentum $\hat b_2$ is driven by the light fluctuations $\hat u_1$ in Eq. (\ref{eqn:b2dot}) \cite{quantum_noise_gardiner}. These fluctuations are transduced by the mechanical oscillator and contribute to the phase quadrature of the output field, 
causing \emph{back action} noise. Since this noise arises from the fluctuations transduced by the mechanical oscillator, it is an instance of a force noise. Contrary to $\nf$, back action cannot be avoided by careful engineering \footnote{Even though it is unavoidable, back action noise can be evaded through specific measurement schemes \cite{qnd_thorne_1978, bae_2014, bae_meas_2019}} 
as it is the measurement noise due to the apparatus (the light field) probing the oscillator \cite{Danilishin_qm_theory}.

The output light responds to the mechanical motion via $\hat b_1$ in the second term of  Eq. (\ref{eqn:v2(t)}). The additive noise $\ns$ to $\hat b_1$ characterizes the difference between the center-of-mass motion and the position that the light actually senses, and is an instance of a sensing noise. This noise could occur because of, for example, the fluctuating coating thickness in mirrors ~\cite{coating_thermal_gras}. We note that in this particular model, the sensing noise is measured (or amplified) according to the interaction strength $\Omq$; which is crucial for the universality of the entangling-disentangling transition. The presence of sensing noises that do not couple to the detection channel at a rate of $\Omq$ prohibit the universality of entanglement. Examples to such noises are detection inefficiency (i.e. photon loss), or additive noise (e.g. dark noise) \cite{quantum_noise_gardiner}.

Lastly, the field of the output mode populated by the vacuum is often called the \emph{shot noise}, which is a sensing noise. Similar to the back action, it is fundamentally related to the measurement process, contrary to $\ns$. For practical examples of environmental noise encountered in an experimental setting, see Sec. II in Ref. \footnotemark[1]. 

\emph{Entanglement Criterion---} We are interested in the optomechanical entanglement between the traveling  light field and the oscillator's motion. 
At any given time $T$, as shown in Fig.~\ref{fig:light_curves}, we have three parties: the oscillator $\{\hat b_{1,2}(T)\}$; the output light field $\{\hat v_{1,2}(t):\,t<T\}$ that {\it has} been reflected during $t<T$ (shown in blue); and input light field $\{\hat u_{1,2}(t):\,t>T\}$ that {\it will} be reflected during $t > T$ (shown in orange). 
Stationary dynamics are invariant under time translations, therefore we set $T=0$.
%

The dynamics in Eqs.~\eqref{eqn:eqns_time_domain} are linear and stable. Since the driving noises are Gaussian, the system reaches a Gaussian steady state completely characterized by the covariance matrix of its quadratures. As we are interested in the bipartite entanglement between the single mechanical mode $ \hat b_j(0)$ and the output field $\{ \hat v_{1,2}(t), t<0\}$, their covariance matrix can be written as the following,
\begin{equation}
\label{eqn:cov_matrix}
\mathbf{V} =  \left[ \begin{array}{cc}
\mathbf{V}^{bb}  & \mathbf{V}^{bv} \\
\mathbf{V}^{vb} & \mathbf{V}^{vv}
\end{array}
\right]=
\left[ \begin{array}{cc}
\langle \hat b_j(0) \hat b_k(0)\rangle_{\rm s}  & \langle \hat b_j(0) \hat v_m(-t')\rangle_{\rm s} \\
\langle \hat v_l(-t) \hat b_k(0)\rangle_{\rm s} & \langle \hat v_l(-t) \hat v_m(-t')\rangle_{\rm s}
\end{array}
\right],
\end{equation}
where $j, k, l, m = 1, 2$, $t >0,\, t' >0$ and we used the symmetrized expectation values
$\langle \hat a \hat b\rangle_{\rm s} \coloneqq \langle\hat a \hat b + \hat b \hat a\rangle/2$. 
More specifically, $\mathbf{V}^{bb}$ is a 2-by-2 matrix of real numbers, $\mathbf{V}^{bv}$ is a block matrix of square-integrable functions on the half real line $\mathcal{L}^2(0, \infty)$, while  $\mathbf{V}^{vv}$ is a block matrix of bounded operators on functions in $\mathcal{L}^2(0, \infty)$.

The mechanical oscillator represents a single mode, while there are $N \to \infty$ countable modes of the output field. In this $1\cross N$ bipartite configuration, the positivity of the partial transpose (PPT) criterion for entanglement is necessary and sufficient \cite{simon_peres-horodecki_2000, werner_bound_2001, adesso_entanglement_2007, braunstein_review, serafini_quantum_2017}. The criterion assesses whether the partially transposed covariance matrix $\mathbf{V}_{\rm{pt}}$ (obtained from $\mathbf{V}$ by taking $\hat b_2 \rightarrow -\hat b_2$) satisfies the Heisenberg uncertainty principle: the system is separable if-and-only-if $\mathbf{V}_{\rm{pt}} + i \mathbf{K}$ is positive semi-definite, with $\mathbf{K} \coloneqq \mathbf{K}^{b} \oplus \mathbf{K}^{v}$\,,  \begin{align}
\label{eqn:commutator_matrix}
\mathbf{K}^b =
\left[\begin{array}{cc}
    0 & 1 \\
    -1 & 0
    \end{array}\right],\;
    \mathbf{K}^v =
     \left[\begin{array}{cc}
    0 & \delta(t) \\
    -\delta(t) & 0
    \end{array}\right] \,,
\end{align}
 the \emph{symplectic form} encoding the commutation relations of the quadratures. In the $1\cross N$ configuration, $\mathbf{V}_{\rm{pt}} + i \mathbf{K}$ can have at most one negative eigenvalue \cite{serafini_quantum_2017}, and we have the following necessary and sufficient test of optomechanical entanglement:
\begin{equation}
\label{eq:charac_poly_def}
    \text{det}\left( \mathbf{V}_\text{pt} + i \mathbf{K} \right) < 0 \iff \text{entanglement}.
\end{equation}

Standard matrix-determinant properties allow to express Eq.~(\ref{eq:charac_poly_def}) as a product of determinants, factoring out $\text{det}\left( \mathbf{V}^{vv} + i\mathbf{K}^v \right)$, which is positive for $\Omq > 0$ \footnote{We disregard the trivial case where $\Omq = 0$: the joint state will be separable, as there is no optomechanical interaction}. We can  then rewrite
\begin{align}
\label{eqn:ent_indicator}
\mathrm{det} \left[ \mathbf{V}^{bb} + i \mathbf{K}^b \right. & \left. - \mathbf{V}^{bv}_{\text{\text{pt}}} \boldsymbol{\cdot} (\mathbf{V}^{vv} + i \mathbf{K}^v)^{-1} \boldsymbol{\cdot} \mathbf{V}^{vb}_{\text{\text{pt}}} \right] < 0 \nonumber \\
& \iff \text{entanglement},
\end{align}
which is the determinant of a $2\times 2$ matrix. Here, we encounter the inversion operation, as well as a \emph{dot product}. When a matrix of operators acts on a matrix of functions, the operators are applied to their respective functions according to the matrix product. 
Due to the dot product being defined only on half of the real line, the inversion operation must be defined causally: for this purpose, we use the Wiener-Hopf method (see Appendix B in \footnotemark[1]).

\emph{Results---}
Let us first assume that the input light field is the vacuum state, or is frequency-independently squeezed.
To observe the nature of entanglement under the effect of environmental noise sources, we fix the shapes of the spectra of the force and the sensing noise to be $S_{n_{\rm{F}}}$ and $S_{n_{\rm{S}}}$, respectively. They are rational functions of the frequency, and can be written in the form of Eq. (\ref{eqn:noise_rational}). We then tune their global amplitudes with parameters $\alpha_{\rm{F}} > 0, \, \beta_{\rm{S}} \geq 0$ \footnote{Note that $\alpha_F$ cannot be zero due to the fluctuation-dissipation theorem.} in the form of $\alpha_{\rm{F}} \, S_{n_{\rm{F}}}$ and $\beta_{\rm{S}} \, S_{n_{\rm{S}}}$. We find that the optomechanical state is:
\begin{enumerate}
    \item separable in the limit of $\beta_{\rm{S}} \rightarrow \infty$ for all $\alpha_{\rm{F}} > 0$,
    \item entangled in the limit of $\beta_{\rm{S}} = 0$, for all $\alpha_{\rm{F}}>0$,
\end{enumerate}
which implies that there exists an entangling-disentangling transition for a finite value of $\beta_{\rm{S}}$. We prove the results above in the companion Article \footnotemark[1], and show that this transition is unique, i.e. it does not occur for multiple values of $\beta_{\rm{S}}$, for a fixed $\alpha_{\rm{F}}$. Therefore, we conclude that in the absence of sensing noise, the optomechanical state is entangled for arbitrary (finite) force noise \footnote{We note that this trivial entanglement in the absence of sensing noise is a potential pitfall. Indeed, models for high-frequency mechanical oscillators typically disregard sensing noise \cite{Aspelmeyer_cavity_optomechanics}. Then, studying entanglement with these models might yield misleading results where entanglement seems very robust, as it is the case in Ref. \cite{gut_stationary_2020}. This point is discussed in detail in Ref. \cite{gut_thesis}.}. 

We propose the following heuristic physical interpretation to explain this phenomenon: entanglement is generated by the two-mode-squeezing interaction included in the optomechanical coupling \cite{Aspelmeyer_cavity_optomechanics}, and two parties---one in a pure Gaussian state, the other in an arbitrary thermal state---become entangled after any two-mode-squeezing interaction \cite{marian_marian}. In the stationary regime, the force noise (and its associated bath) essentially set the state of the mechanical party (similar to thermalization), while the probing light is in a pure Gaussian state. Therefore, it can be expected that the joint optomechanical state after interaction is entangled for any finite force noise. This heuristic argument disregards the possibilities of the mechanical bath and the measuring light being non-Markovian.

\begin{figure*}
\flushleft
\hspace{10mm}
\includegraphics[width=2.\columnwidth]{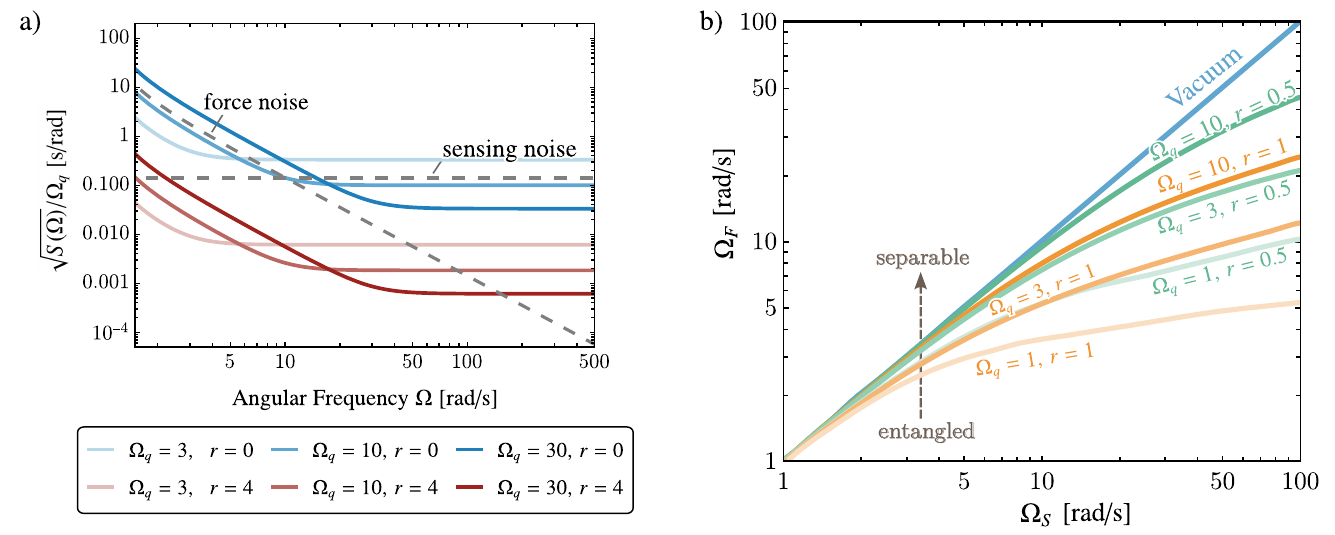}
    \caption{Impact of frequency-dependent squeezing on the entangling-disentangling transition. a) Amplitude  spectra $\sqrt{S(\Omega)}$ normalized by the interaction strength $\Omq$, as a function of different interaction strengths and squeezing levels. Spectra of the force and the sensing noises are shown in dashed lines. Although Markovian at the input, the force noise is transduced by the mechanical oscillator, causing it to decrease as $\Omega^2$ in amplitude  at the output. b) The entangling-disentangling transition with respect to the force and the sensing noises in the system, for different squeezing configurations. The optomechanical state is entangled for the noise configurations below their respective curves. When the input light is in the vacuum state (blue curve), the transition is independent of the interaction strength. We observe that as $\Omq$ increases, the transitions of the other configurations approach that of the vacuum case, in terms of the parameters for which they occur.}
    \label{fig:freq_squeezing_boundary}
\end{figure*}

Simplifying Eq. (\ref{eqn:ent_indicator}), we find that the lhs is independent of $\Omq$, signifying that the {\it entangling-disentangling transition is independent of the coherent optomechanical coupling}. The transition depends solely on the properties of the mechanical oscillator and its environmental noise sources: given an  optomechanical device, if the force and the sensing noises are such that the optomechanical state is separable, increasing the interaction strength $\Omq$ cannot enable the formation of entanglement. 

To understand better the regime where the entangling-disentangling transition takes place, we assume white force and sensing noises, where the double-sided spectra are given by $S_{n_{\rm{F}}}(\Omega) = 2\Omegaf^2/\omegam$ and $S_{n_{\rm{S}}}(\Omega) = 2\omegam/\Omegas^2$, respectively. $\Omegaf$ and $\Omegas$ are the frequencies at which the respective noise spectrum observed at the output field touches the free-mass SQL, defined as $S_\text{SQL}(\Omega) = 2\hbar/M\Omega^2$ \cite{chen_macroscopic_2013} where $M$ is the mass of the oscillator.
Furthermore, we work in the \emph{free-mass limit}, where $\Omegaf, \Omegas \gg \omegam, \gammam$, in order to devise a general formula concerning $\Omegaf$ and $\Omegas$ only \cite{direkci_2024}. We find in \footnotemark[1] that in this limit, the universal transition takes place at $\Omegaf = \Omegas$, as  observed numerically in \cite{direkci_2024}. This corresponds to a total {\it environmental} displacement noise spectrum that is a factor of $2$ away from the free-mass SQL, which  indicates that the free-mass SQL is the relevant scale for characterizing the quantumness of a mechanical oscillator.


%
More generally, suppose that the input light field is prepared with an arbitrary, causal symplectic transformation while maintaining stationarity (in other words, squeezed in a frequency-\emph{dependent} manner). 
In this case,
Eq. (\ref{eqn:ent_indicator}) is no longer independent of $\Omq$, signifying that the entangling-disentangling transition is not universal. 
However, as we show in the companion Article \footnotemark[1], frequency-dependent squeezing does not enhance our ability to create light-mass entanglement.  This is proven in two steps. 
First, we show that in the presence of arbitrary frequency-dependent squeezing, fixing environmental noise levels, entanglement cannot be destroyed by increasing $\Omq$: taking the limit of $\Omq \rightarrow \infty$  is the optimal strategy to generate light-mass  entanglement in that case. We then show that as  $\Omq \rightarrow \infty$, Eq. (\ref{eqn:ent_indicator}) reduces to its form when the input light field is the vacuum state.
Since entanglement is independent of $\Omq$ for the vacuum input, we conclude that for finite $\Omq$, the vacuum input is better than frequency-dependent squeezing for achieving optomechanical entanglement. Intuitively, this result can be understood with the monogamy of entanglement. When the input light field is squeezed in a frequency-dependent fashion, fields that have yet to enter the system (shown in orange in Fig.~\ref{fig:light_curves}) are entangled with the bipartite system consisting of the oscillator (at the origin in Fig.~\ref{fig:light_curves}b) and the reflected light field (shown in blue in Fig.~\ref{fig:light_curves}), constraining the  achievable optomechanical entanglement within this bipartite system. However, we numerically observed that once optomechanical entanglement is achieved, frequency-dependent squeezing can increase or decrease the amount of entanglement depending on the squeezing configuration. Here, entanglement is quantified by an entanglement monotone such as the logarithmic negativity \cite{negativity, plenio_negativity}.




In Fig.~\ref{fig:freq_squeezing_boundary}, we depict example configurations in which frequency-dependent squeezing suppresses the quantum noise spectra $S(\Omega)$ by $e^r$, where $r>0$ is the squeeze factor. This is achieved by first frequency-independently squeezing the light by $e^r$, and then, by filtering it with a detuned cavity \cite{kimble_conversion} (see Appendix G in \footnotemark[1]). We again assume white force and sensing noises, whose spectrum at the phase quadrature of the output light field is plotted in Fig. \ref{fig:freq_squeezing_boundary}a. Note that the force noise spectrum
scales as $\Omega^{-2}$ in amplitude after being transduced by the mechanical oscillator. In Fig. \ref{fig:freq_squeezing_boundary}b, we plot the boundary of the entangling-disentangling transition as a function of the force and sensing noise parameters $\Omegaf$ and $\Omegas$: as mentioned above, the transition occurs for $\Omegaf = \Omegas$ when the input light field is the vacuum state. However, for frequency-dependent squeezing,
the transition is harder to achieve, e.g., it requires a lower sensing noise level (larger $\Omegas$) for a given force noise level $\Omegaf$.
%


\emph{Conclusions---}In this Letter, we showed the existence of a unique and universal entangling-disentangling transition between a mechanical oscillator and the reflected output light field, in the presence of Gaussian environmental noises with arbitrary spectra. 
Then, if the environmental noises are above the transition threshold, one cannot achieve optomechanical entanglement by increasing the interaction strength between the oscillator and the light field. Furthermore, we showed that frequency-dependent squeezing 
is bound to decrease/destroy the amount of optomechanical entanglement in the system. In summary, whether the oscillator is a ``quantum'' (can be entangled with reflected light field) or a ``classical'' (cannot be entangled with reflected light field) object only depends on whether the environmental noise level is below or above the universal transition threshold.

With the techniques developed here, and in \footnotemark[1], experimentalists 
could inquire whether the joint stationary optomechanical state of their device is entangled---in theory. For high frequency devices typically affected by Markovian 
noises, this topic has been discussed theoretically for more than ten years \cite{GENES200933, miao_universal_2010}, and attempts to demonstrate stationary optomechanical entanglement are ongoing \cite{Hoelscher_thesis, mogadas_thesis, gut_thesis}. Providing a quantitative framework to make predictions in the presence of non-Markovian noises is crucial for macroscopic devices operating at low frequencies, since they are typically affected by non-Markovian environments \cite{groblacher_observation_2015}.

Finally, let us consider the fact that before applying {\it coarse graining}, the system that consists of the heat bath, the oscillator and the light field is in a pure state. With knowledge from the bath, are we able to ``reinstate'' optomechanical entanglement?  This is possible in principle, at least in some cases. In our model, $\ns$ can be a classical random process, while $\nf$ can be the superposition of a classical random process and quantum noise that drives the zero-point fluctuations of the oscillator. In principle, one can know the particular realizations of both classical random processes, and hence reduce the environmental noises of the oscillator to its zero-point level---far below the entangling-disentangling transition.

\emph{Acknowledgements---} We thank Aaron Markovitz and Klemens Hammerer for helpful discussions. S.D. and Y.C. acknowledge the support by the Simons Foundation (Award Number 568762), and by the US NSF
grant PHY-2309231. K.W. acknowledges the
support by the Vienna Doctoral School in Physics (VDSP). K.W., C.G., and M.A. received
funding from the European Research Council (ERC) under the
European Union’s Horizon 2020 research and innovation program (Grant Agreement No. 951234), and from the Research
Network Quantum Aspects of Spacetime (TURIS).



\begin{thebibliography}{53}%
\makeatletter
\providecommand \@ifxundefined [1]{%
 \@ifx{#1\undefined}
}%
\providecommand \@ifnum [1]{%
 \ifnum #1\expandafter \@firstoftwo
 \else \expandafter \@secondoftwo
 \fi
}%
\providecommand \@ifx [1]{%
 \ifx #1\expandafter \@firstoftwo
 \else \expandafter \@secondoftwo
 \fi
}%
\providecommand \natexlab [1]{#1}%
\providecommand \enquote  [1]{``#1''}%
\providecommand \bibnamefont  [1]{#1}%
\providecommand \bibfnamefont [1]{#1}%
\providecommand \citenamefont [1]{#1}%
\providecommand \href@noop [0]{\@secondoftwo}%
\providecommand \href [0]{\begingroup \@sanitize@url \@href}%
\providecommand \@href[1]{\@@startlink{#1}\@@href}%
\providecommand \@@href[1]{\endgroup#1\@@endlink}%
\providecommand \@sanitize@url [0]{\catcode `\\12\catcode `\$12\catcode
  `\&12\catcode `\#12\catcode `\^12\catcode `\_12\catcode `\%12\relax}%
\providecommand \@@startlink[1]{}%
\providecommand \@@endlink[0]{}%
\providecommand \url  [0]{\begingroup\@sanitize@url \@url }%
\providecommand \@url [1]{\endgroup\@href {#1}{\urlprefix }}%
\providecommand \urlprefix  [0]{URL }%
\providecommand \Eprint [0]{\href }%
\providecommand \doibase [0]{https://doi.org/}%
\providecommand \selectlanguage [0]{\@gobble}%
\providecommand \bibinfo  [0]{\@secondoftwo}%
\providecommand \bibfield  [0]{\@secondoftwo}%
\providecommand \translation [1]{[#1]}%
\providecommand \BibitemOpen [0]{}%
\providecommand \bibitemStop [0]{}%
\providecommand \bibitemNoStop [0]{.\EOS\space}%
\providecommand \EOS [0]{\spacefactor3000\relax}%
\providecommand \BibitemShut  [1]{\csname bibitem#1\endcsname}%
\let\auto@bib@innerbib\@empty
\bibitem [{\citenamefont {von Neumann}\ and\ \citenamefont
  {Beyer}(2018)}]{Neumann_1955}%
  \BibitemOpen
  \bibfield  {author} {\bibinfo {author} {\bibfnamefont {J.}~\bibnamefont {von
  Neumann}}\ and\ \bibinfo {author} {\bibfnamefont {R.~T.}\ \bibnamefont
  {Beyer}},\ }\href {http://www.jstor.org/stable/j.ctt1wq8zhp} {\emph {\bibinfo
  {title} {Mathematical Foundations of Quantum Mechanics: New Edition}}}\
  (\bibinfo  {publisher} {Princeton University Press},\ \bibinfo {year}
  {2018})\BibitemShut {NoStop}%
\bibitem [{\citenamefont {Zurek}(1981)}]{zurek_pointer}%
  \BibitemOpen
  \bibfield  {author} {\bibinfo {author} {\bibfnamefont {W.~H.}\ \bibnamefont
  {Zurek}},\ }\bibfield  {title} {\bibinfo {title} {Pointer basis of quantum
  apparatus: Into what mixture does the wave packet collapse?},\ }\href
  {https://doi.org/10.1103/PhysRevD.24.1516} {\bibfield  {journal} {\bibinfo
  {journal} {Phys. Rev. D}\ }\textbf {\bibinfo {volume} {24}},\ \bibinfo
  {pages} {1516} (\bibinfo {year} {1981})}\BibitemShut {NoStop}%
\bibitem [{\citenamefont {Joos}\ and\ \citenamefont
  {Zeh}(1985)}]{Joos_Zeh_1985}%
  \BibitemOpen
  \bibfield  {author} {\bibinfo {author} {\bibfnamefont {E.}~\bibnamefont
  {Joos}}\ and\ \bibinfo {author} {\bibfnamefont {H.~D.}\ \bibnamefont {Zeh}},\
  }\bibfield  {title} {\bibinfo {title} {The emergence of classical properties
  through interaction with the environment},\ }\href
  {https://doi.org/10.1007/BF01725541} {\bibfield  {journal} {\bibinfo
  {journal} {Zeitschrift f{\"u}r Physik B Condensed Matter}\ }\textbf {\bibinfo
  {volume} {59}},\ \bibinfo {pages} {223} (\bibinfo {year} {1985})}\BibitemShut
  {NoStop}%
\bibitem [{\citenamefont {Schlosshauer-Selbach}(2008)}]{schlosshauer_book}%
  \BibitemOpen
  \bibfield  {author} {\bibinfo {author} {\bibfnamefont {M.}~\bibnamefont
  {Schlosshauer-Selbach}},\ }\bibfield  {title} {\bibinfo {title} {Decoherence
  and the quantum-to-classical transition}\ }(\bibinfo {year}
  {2008})\BibitemShut {NoStop}%
\bibitem [{\citenamefont {Schlosshauer}(2019)}]{SCHLOSSHAUER2019_decoherence}%
  \BibitemOpen
  \bibfield  {author} {\bibinfo {author} {\bibfnamefont {M.}~\bibnamefont
  {Schlosshauer}},\ }\bibfield  {title} {\bibinfo {title} {Quantum
  decoherence},\ }\href
  {https://doi.org/https://doi.org/10.1016/j.physrep.2019.10.001} {\bibfield
  {journal} {\bibinfo  {journal} {Physics Reports}\ }\textbf {\bibinfo {volume}
  {831}},\ \bibinfo {pages} {1} (\bibinfo {year} {2019})},\ \bibinfo {note}
  {quantum decoherence}\BibitemShut {NoStop}%
\bibitem [{\citenamefont {Genes}\ \emph {et~al.}(2009)\citenamefont {Genes},
  \citenamefont {Mari}, \citenamefont {Vitali},\ and\ \citenamefont
  {Tombesi}}]{GENES200933}%
  \BibitemOpen
  \bibfield  {author} {\bibinfo {author} {\bibfnamefont {C.}~\bibnamefont
  {Genes}}, \bibinfo {author} {\bibfnamefont {A.}~\bibnamefont {Mari}},
  \bibinfo {author} {\bibfnamefont {D.}~\bibnamefont {Vitali}},\ and\ \bibinfo
  {author} {\bibfnamefont {P.}~\bibnamefont {Tombesi}},\ }\bibfield  {title}
  {\bibinfo {title} {Chapter 2 quantum effects in optomechanical systems},\
  }in\ \href {https://doi.org/https://doi.org/10.1016/S1049-250X(09)57002-4}
  {\emph {\bibinfo {booktitle} {Advances in Atomic Molecular and Optical
  Physics}}},\ \bibinfo {series} {Advances In Atomic, Molecular, and Optical
  Physics}, Vol.~\bibinfo {volume} {57}\ (\bibinfo  {publisher} {Academic
  Press},\ \bibinfo {address} {San Diego},\ \bibinfo {year} {2009})\ pp.\
  \bibinfo {pages} {33--86}\BibitemShut {NoStop}%
\bibitem [{\citenamefont {Tang}\ \emph {et~al.}(2022)\citenamefont {Tang},
  \citenamefont {Cai}, \citenamefont {Cheng}, \citenamefont {Xu}, \citenamefont
  {Peng}, \citenamefont {Chen}, \citenamefont {Wang}, \citenamefont {Xia},
  \citenamefont {Wang}, \citenamefont {Song}, \citenamefont {Zhou},\ and\
  \citenamefont {Deng}}]{TANG_perspective_2022}%
  \BibitemOpen
  \bibfield  {author} {\bibinfo {author} {\bibfnamefont {J.-D.}\ \bibnamefont
  {Tang}}, \bibinfo {author} {\bibfnamefont {Q.-Z.}\ \bibnamefont {Cai}},
  \bibinfo {author} {\bibfnamefont {Z.-D.}\ \bibnamefont {Cheng}}, \bibinfo
  {author} {\bibfnamefont {N.}~\bibnamefont {Xu}}, \bibinfo {author}
  {\bibfnamefont {G.-Y.}\ \bibnamefont {Peng}}, \bibinfo {author}
  {\bibfnamefont {P.-Q.}\ \bibnamefont {Chen}}, \bibinfo {author}
  {\bibfnamefont {D.-G.}\ \bibnamefont {Wang}}, \bibinfo {author}
  {\bibfnamefont {Z.-W.}\ \bibnamefont {Xia}}, \bibinfo {author} {\bibfnamefont
  {Y.}~\bibnamefont {Wang}}, \bibinfo {author} {\bibfnamefont {H.-Z.}\
  \bibnamefont {Song}}, \bibinfo {author} {\bibfnamefont {Q.}~\bibnamefont
  {Zhou}},\ and\ \bibinfo {author} {\bibfnamefont {G.-W.}\ \bibnamefont
  {Deng}},\ }\bibfield  {title} {\bibinfo {title} {A perspective on quantum
  entanglement in optomechanical systems},\ }\href
  {https://doi.org/https://doi.org/10.1016/j.physleta.2022.127966} {\bibfield
  {journal} {\bibinfo  {journal} {Physics Letters A}\ }\textbf {\bibinfo
  {volume} {429}},\ \bibinfo {pages} {127966} (\bibinfo {year}
  {2022})}\BibitemShut {NoStop}%
\bibitem [{\citenamefont {Aspelmeyer}\ \emph {et~al.}(2014)\citenamefont
  {Aspelmeyer}, \citenamefont {Kippenberg},\ and\ \citenamefont
  {Marquardt}}]{Aspelmeyer_cavity_optomechanics}%
  \BibitemOpen
  \bibfield  {author} {\bibinfo {author} {\bibfnamefont {M.}~\bibnamefont
  {Aspelmeyer}}, \bibinfo {author} {\bibfnamefont {T.~J.}\ \bibnamefont
  {Kippenberg}},\ and\ \bibinfo {author} {\bibfnamefont {F.}~\bibnamefont
  {Marquardt}},\ }\bibfield  {title} {\bibinfo {title} {Cavity optomechanics},\
  }\href {https://doi.org/10.1103/revmodphys.86.1391} {\bibfield  {journal}
  {\bibinfo  {journal} {Reviews of Modern Physics}\ }\textbf {\bibinfo {volume}
  {86}},\ \bibinfo {pages} {1391} (\bibinfo {year} {2014})}\BibitemShut
  {NoStop}%
\bibitem [{\citenamefont {Hofer}\ \emph {et~al.}(2011)\citenamefont {Hofer},
  \citenamefont {Wieczorek}, \citenamefont {Aspelmeyer},\ and\ \citenamefont
  {Hammerer}}]{hofer_2011}%
  \BibitemOpen
  \bibfield  {author} {\bibinfo {author} {\bibfnamefont {S.~G.}\ \bibnamefont
  {Hofer}}, \bibinfo {author} {\bibfnamefont {W.}~\bibnamefont {Wieczorek}},
  \bibinfo {author} {\bibfnamefont {M.}~\bibnamefont {Aspelmeyer}},\ and\
  \bibinfo {author} {\bibfnamefont {K.}~\bibnamefont {Hammerer}},\ }\bibfield
  {title} {\bibinfo {title} {Quantum entanglement and teleportation in pulsed
  cavity optomechanics},\ }\href {https://doi.org/10.1103/PhysRevA.84.052327}
  {\bibfield  {journal} {\bibinfo  {journal} {Phys. Rev. A}\ }\textbf {\bibinfo
  {volume} {84}},\ \bibinfo {pages} {052327} (\bibinfo {year}
  {2011})}\BibitemShut {NoStop}%
\bibitem [{\citenamefont {Palomaki}\ \emph {et~al.}(2013)\citenamefont
  {Palomaki}, \citenamefont {Teufel}, \citenamefont {Simmonds},\ and\
  \citenamefont {Lehnert}}]{palomaki_2013}%
  \BibitemOpen
  \bibfield  {author} {\bibinfo {author} {\bibfnamefont {T.~A.}\ \bibnamefont
  {Palomaki}}, \bibinfo {author} {\bibfnamefont {J.~D.}\ \bibnamefont
  {Teufel}}, \bibinfo {author} {\bibfnamefont {R.~W.}\ \bibnamefont
  {Simmonds}},\ and\ \bibinfo {author} {\bibfnamefont {K.~W.}\ \bibnamefont
  {Lehnert}},\ }\bibfield  {title} {\bibinfo {title} {Entangling mechanical
  motion with microwave fields},\ }\href
  {https://doi.org/10.1126/science.1244563} {\bibfield  {journal} {\bibinfo
  {journal} {Science}\ }\textbf {\bibinfo {volume} {342}},\ \bibinfo {pages}
  {710} (\bibinfo {year} {2013})},\ \Eprint
  {https://arxiv.org/abs/https://www.science.org/doi/pdf/10.1126/science.1244563}
  {https://www.science.org/doi/pdf/10.1126/science.1244563} \BibitemShut
  {NoStop}%
\bibitem [{\citenamefont {Miao}\ \emph
  {et~al.}(2010{\natexlab{a}})\citenamefont {Miao}, \citenamefont
  {Danilishin},\ and\ \citenamefont {Chen}}]{miao_universal_2010}%
  \BibitemOpen
  \bibfield  {author} {\bibinfo {author} {\bibfnamefont {H.}~\bibnamefont
  {Miao}}, \bibinfo {author} {\bibfnamefont {S.}~\bibnamefont {Danilishin}},\
  and\ \bibinfo {author} {\bibfnamefont {Y.}~\bibnamefont {Chen}},\ }\bibfield
  {title} {\bibinfo {title} {Universal quantum entanglement between an
  oscillator and continuous fields},\ }\href
  {https://doi.org/10.1103/PhysRevA.81.052307} {\bibfield  {journal} {\bibinfo
  {journal} {Physical Review A}\ }\textbf {\bibinfo {volume} {81}},\ \bibinfo
  {pages} {052307} (\bibinfo {year} {2010}{\natexlab{a}})},\ \bibinfo {note}
  {publisher: American Physical Society}\BibitemShut {NoStop}%
\bibitem [{\citenamefont {Miao}\ \emph
  {et~al.}(2010{\natexlab{b}})\citenamefont {Miao}, \citenamefont {Danilishin},
  \citenamefont {M\"uller-Ebhardt}, \citenamefont {Rehbein}, \citenamefont
  {Somiya},\ and\ \citenamefont {Chen}}]{miao_probing_macroscopic}%
  \BibitemOpen
  \bibfield  {author} {\bibinfo {author} {\bibfnamefont {H.}~\bibnamefont
  {Miao}}, \bibinfo {author} {\bibfnamefont {S.}~\bibnamefont {Danilishin}},
  \bibinfo {author} {\bibfnamefont {H.}~\bibnamefont {M\"uller-Ebhardt}},
  \bibinfo {author} {\bibfnamefont {H.}~\bibnamefont {Rehbein}}, \bibinfo
  {author} {\bibfnamefont {K.}~\bibnamefont {Somiya}},\ and\ \bibinfo {author}
  {\bibfnamefont {Y.}~\bibnamefont {Chen}},\ }\bibfield  {title} {\bibinfo
  {title} {Probing macroscopic quantum states with a sub-heisenberg accuracy},\
  }\href {https://doi.org/10.1103/PhysRevA.81.012114} {\bibfield  {journal}
  {\bibinfo  {journal} {Phys. Rev. A}\ }\textbf {\bibinfo {volume} {81}},\
  \bibinfo {pages} {012114} (\bibinfo {year} {2010}{\natexlab{b}})}\BibitemShut
  {NoStop}%
\bibitem [{\citenamefont {Gut}\ \emph {et~al.}(2020)\citenamefont {Gut},
  \citenamefont {Winkler}, \citenamefont {Hoelscher-Obermaier}, \citenamefont
  {Hofer}, \citenamefont {Nia}, \citenamefont {Walk}, \citenamefont {Steffens},
  \citenamefont {Eisert}, \citenamefont {Wieczorek}, \citenamefont {Slater},
  \citenamefont {Aspelmeyer},\ and\ \citenamefont
  {Hammerer}}]{gut_stationary_2020}%
  \BibitemOpen
  \bibfield  {author} {\bibinfo {author} {\bibfnamefont {C.}~\bibnamefont
  {Gut}}, \bibinfo {author} {\bibfnamefont {K.}~\bibnamefont {Winkler}},
  \bibinfo {author} {\bibfnamefont {J.}~\bibnamefont {Hoelscher-Obermaier}},
  \bibinfo {author} {\bibfnamefont {S.~G.}\ \bibnamefont {Hofer}}, \bibinfo
  {author} {\bibfnamefont {R.~M.}\ \bibnamefont {Nia}}, \bibinfo {author}
  {\bibfnamefont {N.}~\bibnamefont {Walk}}, \bibinfo {author} {\bibfnamefont
  {A.}~\bibnamefont {Steffens}}, \bibinfo {author} {\bibfnamefont
  {J.}~\bibnamefont {Eisert}}, \bibinfo {author} {\bibfnamefont
  {W.}~\bibnamefont {Wieczorek}}, \bibinfo {author} {\bibfnamefont {J.~A.}\
  \bibnamefont {Slater}}, \bibinfo {author} {\bibfnamefont {M.}~\bibnamefont
  {Aspelmeyer}},\ and\ \bibinfo {author} {\bibfnamefont {K.}~\bibnamefont
  {Hammerer}},\ }\bibfield  {title} {\bibinfo {title} {Stationary
  optomechanical entanglement between a mechanical oscillator and its
  measurement apparatus},\ }\href
  {https://doi.org/10.1103/PhysRevResearch.2.033244} {\bibfield  {journal}
  {\bibinfo  {journal} {Physical Review Research}\ }\textbf {\bibinfo {volume}
  {2}},\ \bibinfo {pages} {033244} (\bibinfo {year} {2020})}\BibitemShut
  {NoStop}%
\bibitem [{\citenamefont {Direkci}\ \emph {et~al.}(2024)\citenamefont
  {Direkci}, \citenamefont {Winkler}, \citenamefont {Gut}, \citenamefont
  {Hammerer}, \citenamefont {Aspelmeyer},\ and\ \citenamefont
  {Chen}}]{direkci_2024}%
  \BibitemOpen
  \bibfield  {author} {\bibinfo {author} {\bibfnamefont {S.}~\bibnamefont
  {Direkci}}, \bibinfo {author} {\bibfnamefont {K.}~\bibnamefont {Winkler}},
  \bibinfo {author} {\bibfnamefont {C.}~\bibnamefont {Gut}}, \bibinfo {author}
  {\bibfnamefont {K.}~\bibnamefont {Hammerer}}, \bibinfo {author}
  {\bibfnamefont {M.}~\bibnamefont {Aspelmeyer}},\ and\ \bibinfo {author}
  {\bibfnamefont {Y.}~\bibnamefont {Chen}},\ }\bibfield  {title} {\bibinfo
  {title} {Macroscopic quantum entanglement between an optomechanical cavity
  and a continuous field in presence of non-markovian noise},\ }\href
  {https://doi.org/10.1103/PhysRevResearch.6.013175} {\bibfield  {journal}
  {\bibinfo  {journal} {Phys. Rev. Res.}\ }\textbf {\bibinfo {volume} {6}},\
  \bibinfo {pages} {013175} (\bibinfo {year} {2024})}\BibitemShut {NoStop}%
\bibitem [{\citenamefont {Paternostro}\ \emph {et~al.}(2007)\citenamefont
  {Paternostro}, \citenamefont {Vitali}, \citenamefont {Gigan}, \citenamefont
  {Kim}, \citenamefont {Brukner}, \citenamefont {Eisert},\ and\ \citenamefont
  {Aspelmeyer}}]{PhysRevLett.99.250401}%
  \BibitemOpen
  \bibfield  {author} {\bibinfo {author} {\bibfnamefont {M.}~\bibnamefont
  {Paternostro}}, \bibinfo {author} {\bibfnamefont {D.}~\bibnamefont {Vitali}},
  \bibinfo {author} {\bibfnamefont {S.}~\bibnamefont {Gigan}}, \bibinfo
  {author} {\bibfnamefont {M.~S.}\ \bibnamefont {Kim}}, \bibinfo {author}
  {\bibfnamefont {C.}~\bibnamefont {Brukner}}, \bibinfo {author} {\bibfnamefont
  {J.}~\bibnamefont {Eisert}},\ and\ \bibinfo {author} {\bibfnamefont
  {M.}~\bibnamefont {Aspelmeyer}},\ }\bibfield  {title} {\bibinfo {title}
  {Creating and probing multipartite macroscopic entanglement with light},\
  }\href {https://doi.org/10.1103/PhysRevLett.99.250401} {\bibfield  {journal}
  {\bibinfo  {journal} {Phys. Rev. Lett.}\ }\textbf {\bibinfo {volume} {99}},\
  \bibinfo {pages} {250401} (\bibinfo {year} {2007})}\BibitemShut {NoStop}%
\bibitem [{\citenamefont {Vitali}\ \emph {et~al.}(2007)\citenamefont {Vitali},
  \citenamefont {Gigan}, \citenamefont {Ferreira}, \citenamefont {B\"ohm},
  \citenamefont {Tombesi}, \citenamefont {Guerreiro}, \citenamefont {Vedral},
  \citenamefont {Zeilinger},\ and\ \citenamefont
  {Aspelmeyer}}]{PhysRevLett.98.030405}%
  \BibitemOpen
  \bibfield  {author} {\bibinfo {author} {\bibfnamefont {D.}~\bibnamefont
  {Vitali}}, \bibinfo {author} {\bibfnamefont {S.}~\bibnamefont {Gigan}},
  \bibinfo {author} {\bibfnamefont {A.}~\bibnamefont {Ferreira}}, \bibinfo
  {author} {\bibfnamefont {H.~R.}\ \bibnamefont {B\"ohm}}, \bibinfo {author}
  {\bibfnamefont {P.}~\bibnamefont {Tombesi}}, \bibinfo {author} {\bibfnamefont
  {A.}~\bibnamefont {Guerreiro}}, \bibinfo {author} {\bibfnamefont
  {V.}~\bibnamefont {Vedral}}, \bibinfo {author} {\bibfnamefont
  {A.}~\bibnamefont {Zeilinger}},\ and\ \bibinfo {author} {\bibfnamefont
  {M.}~\bibnamefont {Aspelmeyer}},\ }\bibfield  {title} {\bibinfo {title}
  {Optomechanical entanglement between a movable mirror and a cavity field},\
  }\href {https://doi.org/10.1103/PhysRevLett.98.030405} {\bibfield  {journal}
  {\bibinfo  {journal} {Phys. Rev. Lett.}\ }\textbf {\bibinfo {volume} {98}},\
  \bibinfo {pages} {030405} (\bibinfo {year} {2007})}\BibitemShut {NoStop}%
\bibitem [{Note1()}]{Note1}%
  \BibitemOpen
  \bibinfo {note} {See the companion Article titled ``Characterizing Stationary
  Optomechanical Entanglement in the Presence of Non-Markovian
  Noise''.}\BibitemShut {Stop}%
\bibitem [{\citenamefont {Danilishin}\ and\ \citenamefont
  {Khalili}(2012)}]{Danilishin_qm_theory}%
  \BibitemOpen
  \bibfield  {author} {\bibinfo {author} {\bibfnamefont {S.~L.}\ \bibnamefont
  {Danilishin}}\ and\ \bibinfo {author} {\bibfnamefont {F.~Y.}\ \bibnamefont
  {Khalili}},\ }\bibfield  {title} {\bibinfo {title} {Quantum measurement
  theory in gravitational-wave detectors},\ }\href
  {https://doi.org/10.12942/lrr-2012-5} {\bibfield  {journal} {\bibinfo
  {journal} {Living Reviews in Relativity}\ }\textbf {\bibinfo {volume} {15}},\
  \bibinfo {pages} {5} (\bibinfo {year} {2012})}\BibitemShut {NoStop}%
\bibitem [{\citenamefont {Chen}(2013)}]{chen_macroscopic_2013}%
  \BibitemOpen
  \bibfield  {author} {\bibinfo {author} {\bibfnamefont {Y.}~\bibnamefont
  {Chen}},\ }\bibfield  {title} {\bibinfo {title} {Macroscopic quantum
  mechanics: theory and experimental concepts of optomechanics},\ }\href
  {https://doi.org/10.1088/0953-4075/46/10/104001} {\bibfield  {journal}
  {\bibinfo  {journal} {Journal of Physics B: Atomic, Molecular and Optical
  Physics}\ }\textbf {\bibinfo {volume} {46}},\ \bibinfo {pages} {104001}
  (\bibinfo {year} {2013})},\ \bibinfo {note} {publisher: IOP
  Publishing}\BibitemShut {NoStop}%
\bibitem [{\citenamefont {Collaboration}(2015)}]{advanced_ligo}%
  \BibitemOpen
  \bibfield  {author} {\bibinfo {author} {\bibfnamefont {T.~L.~S.}\
  \bibnamefont {Collaboration}},\ }\bibfield  {title} {\bibinfo {title}
  {Advanced ligo},\ }\href {https://doi.org/10.1088/0264-9381/32/7/074001}
  {\bibfield  {journal} {\bibinfo  {journal} {Classical and Quantum Gravity}\
  }\textbf {\bibinfo {volume} {32}},\ \bibinfo {pages} {074001} (\bibinfo
  {year} {2015})}\BibitemShut {NoStop}%
\bibitem [{\citenamefont {Yu}\ \emph {et~al.}(2020)\citenamefont {Yu},
  \citenamefont {McCuller}, \citenamefont {Tse}, \citenamefont {Kijbunchoo},
  \citenamefont {Barsotti},\ and\ \citenamefont {Mavalvala}}]{yu2020quantum}%
  \BibitemOpen
  \bibfield  {author} {\bibinfo {author} {\bibfnamefont {H.}~\bibnamefont
  {Yu}}, \bibinfo {author} {\bibfnamefont {L.}~\bibnamefont {McCuller}},
  \bibinfo {author} {\bibfnamefont {M.}~\bibnamefont {Tse}}, \bibinfo {author}
  {\bibfnamefont {N.}~\bibnamefont {Kijbunchoo}}, \bibinfo {author}
  {\bibfnamefont {L.}~\bibnamefont {Barsotti}},\ and\ \bibinfo {author}
  {\bibfnamefont {N.}~\bibnamefont {Mavalvala}},\ }\bibfield  {title} {\bibinfo
  {title} {Quantum correlations between light and the kilogram-mass mirrors of
  ligo},\ }\href@noop {} {\bibfield  {journal} {\bibinfo  {journal} {Nature}\
  }\textbf {\bibinfo {volume} {583}},\ \bibinfo {pages} {43} (\bibinfo {year}
  {2020})}\BibitemShut {NoStop}%
\bibitem [{\citenamefont {McCuller}\ \emph {et~al.}(2020)\citenamefont
  {McCuller}, \citenamefont {Whittle}, \citenamefont {Ganapathy}, \citenamefont
  {Komori}, \citenamefont {Tse}, \citenamefont {Fernandez-Galiana},
  \citenamefont {Barsotti}, \citenamefont {Fritschel}, \citenamefont
  {MacInnis}, \citenamefont {Matichard}, \citenamefont {Mason}, \citenamefont
  {Mavalvala}, \citenamefont {Mittleman}, \citenamefont {Yu}, \citenamefont
  {Zucker},\ and\ \citenamefont {Evans}}]{lee_ligo_freq_squeezing}%
  \BibitemOpen
  \bibfield  {author} {\bibinfo {author} {\bibfnamefont {L.}~\bibnamefont
  {McCuller}}, \bibinfo {author} {\bibfnamefont {C.}~\bibnamefont {Whittle}},
  \bibinfo {author} {\bibfnamefont {D.}~\bibnamefont {Ganapathy}}, \bibinfo
  {author} {\bibfnamefont {K.}~\bibnamefont {Komori}}, \bibinfo {author}
  {\bibfnamefont {M.}~\bibnamefont {Tse}}, \bibinfo {author} {\bibfnamefont
  {A.}~\bibnamefont {Fernandez-Galiana}}, \bibinfo {author} {\bibfnamefont
  {L.}~\bibnamefont {Barsotti}}, \bibinfo {author} {\bibfnamefont
  {P.}~\bibnamefont {Fritschel}}, \bibinfo {author} {\bibfnamefont
  {M.}~\bibnamefont {MacInnis}}, \bibinfo {author} {\bibfnamefont
  {F.}~\bibnamefont {Matichard}}, \bibinfo {author} {\bibfnamefont
  {K.}~\bibnamefont {Mason}}, \bibinfo {author} {\bibfnamefont
  {N.}~\bibnamefont {Mavalvala}}, \bibinfo {author} {\bibfnamefont
  {R.}~\bibnamefont {Mittleman}}, \bibinfo {author} {\bibfnamefont
  {H.}~\bibnamefont {Yu}}, \bibinfo {author} {\bibfnamefont {M.~E.}\
  \bibnamefont {Zucker}},\ and\ \bibinfo {author} {\bibfnamefont
  {M.}~\bibnamefont {Evans}},\ }\bibfield  {title} {\bibinfo {title}
  {Frequency-dependent squeezing for advanced ligo},\ }\href
  {https://doi.org/10.1103/PhysRevLett.124.171102} {\bibfield  {journal}
  {\bibinfo  {journal} {Phys. Rev. Lett.}\ }\textbf {\bibinfo {volume} {124}},\
  \bibinfo {pages} {171102} (\bibinfo {year} {2020})}\BibitemShut {NoStop}%
\bibitem [{\citenamefont {Ho}\ \emph {et~al.}(2018)\citenamefont {Ho},
  \citenamefont {Oudot}, \citenamefont {Bancal},\ and\ \citenamefont
  {Sangouard}}]{optomech_entanglement_photon_counting}%
  \BibitemOpen
  \bibfield  {author} {\bibinfo {author} {\bibfnamefont {M.}~\bibnamefont
  {Ho}}, \bibinfo {author} {\bibfnamefont {E.}~\bibnamefont {Oudot}}, \bibinfo
  {author} {\bibfnamefont {J.-D.}\ \bibnamefont {Bancal}},\ and\ \bibinfo
  {author} {\bibfnamefont {N.}~\bibnamefont {Sangouard}},\ }\bibfield  {title}
  {\bibinfo {title} {Witnessing optomechanical entanglement with photon
  counting},\ }\href {https://doi.org/10.1103/PhysRevLett.121.023602}
  {\bibfield  {journal} {\bibinfo  {journal} {Phys. Rev. Lett.}\ }\textbf
  {\bibinfo {volume} {121}},\ \bibinfo {pages} {023602} (\bibinfo {year}
  {2018})}\BibitemShut {NoStop}%
\bibitem [{Note2()}]{Note2}%
  \BibitemOpen
  \bibinfo {note} {The factor of $2$ is a particular normalization that
  simplifies the derivations, and it is found in several works with continuous
  variables, see \cite {cv_qi_adesso} for example}\BibitemShut {NoStop}%
\bibitem [{\citenamefont {M\"uller-Ebhardt}\ \emph {et~al.}(2008)\citenamefont
  {M\"uller-Ebhardt}, \citenamefont {Rehbein}, \citenamefont {Schnabel},
  \citenamefont {Danzmann},\ and\ \citenamefont {Chen}}]{Muller-Ebhardt_2008}%
  \BibitemOpen
  \bibfield  {author} {\bibinfo {author} {\bibfnamefont {H.}~\bibnamefont
  {M\"uller-Ebhardt}}, \bibinfo {author} {\bibfnamefont {H.}~\bibnamefont
  {Rehbein}}, \bibinfo {author} {\bibfnamefont {R.}~\bibnamefont {Schnabel}},
  \bibinfo {author} {\bibfnamefont {K.}~\bibnamefont {Danzmann}},\ and\
  \bibinfo {author} {\bibfnamefont {Y.}~\bibnamefont {Chen}},\ }\bibfield
  {title} {\bibinfo {title} {Entanglement of macroscopic test masses and the
  standard quantum limit in laser interferometry},\ }\href
  {https://doi.org/10.1103/PhysRevLett.100.013601} {\bibfield  {journal}
  {\bibinfo  {journal} {Phys. Rev. Lett.}\ }\textbf {\bibinfo {volume} {100}},\
  \bibinfo {pages} {013601} (\bibinfo {year} {2008})}\BibitemShut {NoStop}%
\bibitem [{\citenamefont {Hofer}(2015)}]{hofer_thesis}%
  \BibitemOpen
  \bibfield  {author} {\bibinfo {author} {\bibfnamefont {S.}~\bibnamefont
  {Hofer}},\ }\emph {\bibinfo {title} {Quantum Control of Optomechanical
  Systems}},\ \href {https://doi.org/10.25365/thesis.38975} {\bibinfo {type}
  {Phd thesis}},\ \bibinfo  {school} {University of Vienna}, \bibinfo {address}
  {Vienna, Austria} (\bibinfo {year} {2015})\BibitemShut {NoStop}%
\bibitem [{\citenamefont {Adesso}\ and\ \citenamefont
  {Illuminati}(2007)}]{adesso_entanglement_2007}%
  \BibitemOpen
  \bibfield  {author} {\bibinfo {author} {\bibfnamefont {G.}~\bibnamefont
  {Adesso}}\ and\ \bibinfo {author} {\bibfnamefont {F.}~\bibnamefont
  {Illuminati}},\ }\bibfield  {title} {\bibinfo {title} {Entanglement in
  continuous variable systems: {Recent} advances and current perspectives},\
  }\href {https://doi.org/10.1088/1751-8113/40/28/S01} {\bibfield  {journal}
  {\bibinfo  {journal} {Journal of Physics A: Mathematical and Theoretical}\
  }\textbf {\bibinfo {volume} {40}},\ \bibinfo {pages} {7821} (\bibinfo {year}
  {2007})},\ \bibinfo {note} {arXiv:quant-ph/0701221}\BibitemShut {NoStop}%
\bibitem [{Note3()}]{Note3}%
  \BibitemOpen
  \bibinfo {note} {In terms of system parameters, $\protect \ensuremath {\Omega
  _{\protect \rm {q}}}= \protect \sqrt {\protect \frac {I_{\protect \rm
  {probe}} \omega _{\protect \rm {probe}} }{M c^2}}$, where $I_{\protect \rm
  {probe}}$ and $\omega _{\protect \rm {probe}}$ are the intensity and the
  frequency of the light probing the mechanics, $c$ is the speed of light in
  vacuum, and $M$ is the mass of the oscillator.}\BibitemShut {Stop}%
\bibitem [{\citenamefont {Kubo}(1966)}]{Kubo_1966}%
  \BibitemOpen
  \bibfield  {author} {\bibinfo {author} {\bibfnamefont {R.}~\bibnamefont
  {Kubo}},\ }\bibfield  {title} {\bibinfo {title} {The fluctuation-dissipation
  theorem},\ }\href {https://doi.org/10.1088/0034-4885/29/1/306} {\bibfield
  {journal} {\bibinfo  {journal} {Reports on Progress in Physics}\ }\textbf
  {\bibinfo {volume} {29}},\ \bibinfo {pages} {255} (\bibinfo {year}
  {1966})}\BibitemShut {NoStop}%
\bibitem [{\citenamefont {Callen}\ and\ \citenamefont
  {Welton}(1951)}]{callen_1951}%
  \BibitemOpen
  \bibfield  {author} {\bibinfo {author} {\bibfnamefont {H.~B.}\ \bibnamefont
  {Callen}}\ and\ \bibinfo {author} {\bibfnamefont {T.~A.}\ \bibnamefont
  {Welton}},\ }\bibfield  {title} {\bibinfo {title} {Irreversibility and
  generalized noise},\ }\href {https://doi.org/10.1103/PhysRev.83.34}
  {\bibfield  {journal} {\bibinfo  {journal} {Phys. Rev.}\ }\textbf {\bibinfo
  {volume} {83}},\ \bibinfo {pages} {34} (\bibinfo {year} {1951})}\BibitemShut
  {NoStop}%
\bibitem [{\citenamefont {Wiseman}\ and\ \citenamefont
  {Milburn}(2009)}]{Wiseman_Milburn_control}%
  \BibitemOpen
  \bibfield  {author} {\bibinfo {author} {\bibfnamefont {H.~M.}\ \bibnamefont
  {Wiseman}}\ and\ \bibinfo {author} {\bibfnamefont {G.~J.}\ \bibnamefont
  {Milburn}},\ }\href@noop {} {\emph {\bibinfo {title} {Quantum Measurement and
  Control}}}\ (\bibinfo  {publisher} {Cambridge University Press},\ \bibinfo
  {year} {2009})\BibitemShut {NoStop}%
\bibitem [{\citenamefont {Gardiner}\ and\ \citenamefont
  {Zoller}(2004)}]{quantum_noise_gardiner}%
  \BibitemOpen
  \bibfield  {author} {\bibinfo {author} {\bibfnamefont {C.~W.}\ \bibnamefont
  {Gardiner}}\ and\ \bibinfo {author} {\bibfnamefont {P.}~\bibnamefont
  {Zoller}},\ }\href@noop {} {\emph {\bibinfo {title} {Quantum noise : a
  handbook of Markovian and non-Markovian quantum stochastic methods with
  applications to quantum optics}}},\ \bibinfo {edition} {3rd}\ ed.,\ Springer
  complexity\ (\bibinfo  {publisher} {Springer},\ \bibinfo {address} {Berlin},\
  \bibinfo {year} {2004})\BibitemShut {NoStop}%
\bibitem [{Note4()}]{Note4}%
  \BibitemOpen
  \bibinfo {note} {Even though it is unavoidable, back action noise can be
  evaded through specific measurement schemes \cite {qnd_thorne_1978, bae_2014,
  bae_meas_2019}}\BibitemShut {NoStop}%
\bibitem [{\citenamefont {Gras}\ and\ \citenamefont
  {Evans}(2018)}]{coating_thermal_gras}%
  \BibitemOpen
  \bibfield  {author} {\bibinfo {author} {\bibfnamefont {S.}~\bibnamefont
  {Gras}}\ and\ \bibinfo {author} {\bibfnamefont {M.}~\bibnamefont {Evans}},\
  }\bibfield  {title} {\bibinfo {title} {Direct measurement of coating thermal
  noise in optical resonators},\ }\href
  {https://doi.org/10.1103/PhysRevD.98.122001} {\bibfield  {journal} {\bibinfo
  {journal} {Phys. Rev. D}\ }\textbf {\bibinfo {volume} {98}},\ \bibinfo
  {pages} {122001} (\bibinfo {year} {2018})}\BibitemShut {NoStop}%
\bibitem [{\citenamefont {Simon}(2000)}]{simon_peres-horodecki_2000}%
  \BibitemOpen
  \bibfield  {author} {\bibinfo {author} {\bibfnamefont {R.}~\bibnamefont
  {Simon}},\ }\bibfield  {title} {\bibinfo {title} {Peres-{Horodecki}
  {Separability} {Criterion} for {Continuous} {Variable} {Systems}},\ }\href
  {https://doi.org/10.1103/PhysRevLett.84.2726} {\bibfield  {journal} {\bibinfo
   {journal} {Physical Review Letters}\ }\textbf {\bibinfo {volume} {84}},\
  \bibinfo {pages} {2726} (\bibinfo {year} {2000})}\BibitemShut {NoStop}%
\bibitem [{\citenamefont {Werner}\ and\ \citenamefont
  {Wolf}(2001)}]{werner_bound_2001}%
  \BibitemOpen
  \bibfield  {author} {\bibinfo {author} {\bibfnamefont {R.~F.}\ \bibnamefont
  {Werner}}\ and\ \bibinfo {author} {\bibfnamefont {M.~M.}\ \bibnamefont
  {Wolf}},\ }\bibfield  {title} {\bibinfo {title} {Bound {Entangled} {Gaussian}
  {States}},\ }\href {https://doi.org/10.1103/PhysRevLett.86.3658} {\bibfield
  {journal} {\bibinfo  {journal} {Physical Review Letters}\ }\textbf {\bibinfo
  {volume} {86}},\ \bibinfo {pages} {3658} (\bibinfo {year}
  {2001})}\BibitemShut {NoStop}%
\bibitem [{\citenamefont {Braunstein}\ and\ \citenamefont {van
  Loock}(2005)}]{braunstein_review}%
  \BibitemOpen
  \bibfield  {author} {\bibinfo {author} {\bibfnamefont {S.~L.}\ \bibnamefont
  {Braunstein}}\ and\ \bibinfo {author} {\bibfnamefont {P.}~\bibnamefont {van
  Loock}},\ }\bibfield  {title} {\bibinfo {title} {Quantum information with
  continuous variables},\ }\href {https://doi.org/10.1103/RevModPhys.77.513}
  {\bibfield  {journal} {\bibinfo  {journal} {Rev. Mod. Phys.}\ }\textbf
  {\bibinfo {volume} {77}},\ \bibinfo {pages} {513} (\bibinfo {year}
  {2005})}\BibitemShut {NoStop}%
\bibitem [{\citenamefont {Serafini}(2017)}]{serafini_quantum_2017}%
  \BibitemOpen
  \bibfield  {author} {\bibinfo {author} {\bibfnamefont {A.}~\bibnamefont
  {Serafini}},\ }\href@noop {} {\emph {\bibinfo {title} {Quantum continuous
  variables: a primer of theoretical methods}}}\ (\bibinfo  {publisher} {CRC
  Press, Taylor \& Francis Group, CRC Press is an imprint of the Taylor \&
  Francis Group, an informa business},\ \bibinfo {address} {Boca Raton},\
  \bibinfo {year} {2017})\BibitemShut {NoStop}%
\bibitem [{Note5()}]{Note5}%
  \BibitemOpen
  \bibinfo {note} {We disregard the trivial case where $\protect \ensuremath
  {\Omega _{\protect \rm {q}}}= 0$: the joint state will be separable, as there
  is no optomechanical interaction}\BibitemShut {NoStop}%
\bibitem [{Note6()}]{Note6}%
  \BibitemOpen
  \bibinfo {note} {Note that $\alpha _F$ cannot be zero due to the
  fluctuation-dissipation theorem.}\BibitemShut {Stop}%
\bibitem [{Note7()}]{Note7}%
  \BibitemOpen
  \bibinfo {note} {We note that this trivial entanglement in the absence of
  sensing noise is a potential pitfall. Indeed, models for high-frequency
  mechanical oscillators typically disregard sensing noise \cite
  {Aspelmeyer_cavity_optomechanics}. Then, studying entanglement with these
  models might yield misleading results where entanglement seems very robust,
  as it is the case in Ref. \cite {gut_stationary_2020}. This point is
  discussed in detail in Ref. \cite {gut_thesis}.}\BibitemShut {Stop}%
\bibitem [{\citenamefont {Marian}\ and\ \citenamefont
  {Marian}(2008)}]{marian_marian}%
  \BibitemOpen
  \bibfield  {author} {\bibinfo {author} {\bibfnamefont {P.}~\bibnamefont
  {Marian}}\ and\ \bibinfo {author} {\bibfnamefont {T.~A.}\ \bibnamefont
  {Marian}},\ }\bibfield  {title} {\bibinfo {title} {Entanglement of formation
  for an arbitrary two-mode gaussian state},\ }\href
  {https://doi.org/10.1103/PhysRevLett.101.220403} {\bibfield  {journal}
  {\bibinfo  {journal} {Phys. Rev. Lett.}\ }\textbf {\bibinfo {volume} {101}},\
  \bibinfo {pages} {220403} (\bibinfo {year} {2008})}\BibitemShut {NoStop}%
\bibitem [{\citenamefont {Vidal}\ and\ \citenamefont
  {Werner}(2002)}]{negativity}%
  \BibitemOpen
  \bibfield  {author} {\bibinfo {author} {\bibfnamefont {G.}~\bibnamefont
  {Vidal}}\ and\ \bibinfo {author} {\bibfnamefont {R.~F.}\ \bibnamefont
  {Werner}},\ }\bibfield  {title} {\bibinfo {title} {Computable measure of
  entanglement},\ }\href {https://doi.org/10.1103/PhysRevA.65.032314}
  {\bibfield  {journal} {\bibinfo  {journal} {Phys. Rev. A}\ }\textbf {\bibinfo
  {volume} {65}},\ \bibinfo {pages} {032314} (\bibinfo {year}
  {2002})}\BibitemShut {NoStop}%
\bibitem [{\citenamefont {Plenio}(2005)}]{plenio_negativity}%
  \BibitemOpen
  \bibfield  {author} {\bibinfo {author} {\bibfnamefont {M.~B.}\ \bibnamefont
  {Plenio}},\ }\bibfield  {title} {\bibinfo {title} {Logarithmic negativity: A
  full entanglement monotone that is not convex},\ }\href
  {https://doi.org/10.1103/PhysRevLett.95.090503} {\bibfield  {journal}
  {\bibinfo  {journal} {Phys. Rev. Lett.}\ }\textbf {\bibinfo {volume} {95}},\
  \bibinfo {pages} {090503} (\bibinfo {year} {2005})}\BibitemShut {NoStop}%
\bibitem [{\citenamefont {Kimble}\ \emph {et~al.}(2001)\citenamefont {Kimble},
  \citenamefont {Levin}, \citenamefont {Matsko}, \citenamefont {Thorne},\ and\
  \citenamefont {Vyatchanin}}]{kimble_conversion}%
  \BibitemOpen
  \bibfield  {author} {\bibinfo {author} {\bibfnamefont {H.~J.}\ \bibnamefont
  {Kimble}}, \bibinfo {author} {\bibfnamefont {Y.}~\bibnamefont {Levin}},
  \bibinfo {author} {\bibfnamefont {A.~B.}\ \bibnamefont {Matsko}}, \bibinfo
  {author} {\bibfnamefont {K.~S.}\ \bibnamefont {Thorne}},\ and\ \bibinfo
  {author} {\bibfnamefont {S.~P.}\ \bibnamefont {Vyatchanin}},\ }\bibfield
  {title} {\bibinfo {title} {Conversion of conventional gravitational-wave
  interferometers into quantum nondemolition interferometers by modifying their
  input and/or output optics},\ }\href
  {https://doi.org/10.1103/PhysRevD.65.022002} {\bibfield  {journal} {\bibinfo
  {journal} {Phys. Rev. D}\ }\textbf {\bibinfo {volume} {65}},\ \bibinfo
  {pages} {022002} (\bibinfo {year} {2001})}\BibitemShut {NoStop}%
\bibitem [{\citenamefont {Hoelscher}(2017)}]{Hoelscher_thesis}%
  \BibitemOpen
  \bibfield  {author} {\bibinfo {author} {\bibfnamefont {J.}~\bibnamefont
  {Hoelscher}},\ }\emph {\bibinfo {title} {Generation and Detection of Quantum
  Entanglement in Optomechanical Systems}},\ \href@noop {} {Ph.D. thesis},\
  \bibinfo  {school} {University of Vienna} (\bibinfo {year}
  {2017})\BibitemShut {NoStop}%
\bibitem [{\citenamefont {Nia}(2018)}]{mogadas_thesis}%
  \BibitemOpen
  \bibfield  {author} {\bibinfo {author} {\bibfnamefont {R.~M.}\ \bibnamefont
  {Nia}},\ }\emph {\bibinfo {title} {Multimode optomechanics in the strong
  cooperativity regime : towards optomechanical entanglement with
  micromechanical membranes}},\ \href@noop {} {Ph.D. thesis},\ \bibinfo
  {school} {University of Vienna} (\bibinfo {year} {2018})\BibitemShut
  {NoStop}%
\bibitem [{\citenamefont {Gut}(2024)}]{gut_thesis}%
  \BibitemOpen
  \bibfield  {author} {\bibinfo {author} {\bibfnamefont {C.}~\bibnamefont
  {Gut}},\ }\emph {\bibinfo {title} {Stationary optomechanical entanglement
  detection}},\ \href {https://doi.org/10.25365/thesis.77115} {\bibinfo {type}
  {Phd thesis}},\ \bibinfo  {school} {University of Vienna}, \bibinfo {address}
  {Vienna, Austria} (\bibinfo {year} {2024})\BibitemShut {NoStop}%
\bibitem [{\citenamefont {Gröblacher}\ \emph {et~al.}(2015)\citenamefont
  {Gröblacher}, \citenamefont {Trubarov}, \citenamefont {Prigge},
  \citenamefont {Cole}, \citenamefont {Aspelmeyer},\ and\ \citenamefont
  {Eisert}}]{groblacher_observation_2015}%
  \BibitemOpen
  \bibfield  {author} {\bibinfo {author} {\bibfnamefont {S.}~\bibnamefont
  {Gröblacher}}, \bibinfo {author} {\bibfnamefont {A.}~\bibnamefont
  {Trubarov}}, \bibinfo {author} {\bibfnamefont {N.}~\bibnamefont {Prigge}},
  \bibinfo {author} {\bibfnamefont {G.~D.}\ \bibnamefont {Cole}}, \bibinfo
  {author} {\bibfnamefont {M.}~\bibnamefont {Aspelmeyer}},\ and\ \bibinfo
  {author} {\bibfnamefont {J.}~\bibnamefont {Eisert}},\ }\bibfield  {title}
  {\bibinfo {title} {Observation of non-{Markovian} micromechanical {Brownian}
  motion},\ }\href {https://doi.org/10.1038/ncomms8606} {\bibfield  {journal}
  {\bibinfo  {journal} {Nature Communications}\ }\textbf {\bibinfo {volume}
  {6}},\ \bibinfo {pages} {7606} (\bibinfo {year} {2015})}\BibitemShut
  {NoStop}%
\bibitem [{\citenamefont {Adesso}\ \emph {et~al.}(2014)\citenamefont {Adesso},
  \citenamefont {Ragy},\ and\ \citenamefont {Lee}}]{cv_qi_adesso}%
  \BibitemOpen
  \bibfield  {author} {\bibinfo {author} {\bibfnamefont {G.}~\bibnamefont
  {Adesso}}, \bibinfo {author} {\bibfnamefont {S.}~\bibnamefont {Ragy}},\ and\
  \bibinfo {author} {\bibfnamefont {A.~R.}\ \bibnamefont {Lee}},\ }\bibfield
  {title} {\bibinfo {title} {Continuous variable quantum information: Gaussian
  states and beyond},\ }\href {https://doi.org/10.1142/S1230161214400010}
  {\bibfield  {journal} {\bibinfo  {journal} {Open Systems \& Information
  Dynamics}\ }\textbf {\bibinfo {volume} {21}},\ \bibinfo {pages} {1440001}
  (\bibinfo {year} {2014})},\ \Eprint
  {https://arxiv.org/abs/https://doi.org/10.1142/S1230161214400010}
  {https://doi.org/10.1142/S1230161214400010} \BibitemShut {NoStop}%
\bibitem [{\citenamefont {Thorne}\ \emph {et~al.}(1978)\citenamefont {Thorne},
  \citenamefont {Drever}, \citenamefont {Caves}, \citenamefont {Zimmermann},\
  and\ \citenamefont {Sandberg}}]{qnd_thorne_1978}%
  \BibitemOpen
  \bibfield  {author} {\bibinfo {author} {\bibfnamefont {K.~S.}\ \bibnamefont
  {Thorne}}, \bibinfo {author} {\bibfnamefont {R.~W.~P.}\ \bibnamefont
  {Drever}}, \bibinfo {author} {\bibfnamefont {C.~M.}\ \bibnamefont {Caves}},
  \bibinfo {author} {\bibfnamefont {M.}~\bibnamefont {Zimmermann}},\ and\
  \bibinfo {author} {\bibfnamefont {V.~D.}\ \bibnamefont {Sandberg}},\
  }\bibfield  {title} {\bibinfo {title} {Quantum nondemolition measurements of
  harmonic oscillators},\ }\href {https://doi.org/10.1103/PhysRevLett.40.667}
  {\bibfield  {journal} {\bibinfo  {journal} {Phys. Rev. Lett.}\ }\textbf
  {\bibinfo {volume} {40}},\ \bibinfo {pages} {667} (\bibinfo {year}
  {1978})}\BibitemShut {NoStop}%
\bibitem [{\citenamefont {Suh}\ \emph {et~al.}(2014)\citenamefont {Suh},
  \citenamefont {Weinstein}, \citenamefont {Lei}, \citenamefont {Wollman},
  \citenamefont {Steinke}, \citenamefont {Meystre}, \citenamefont {Clerk},\
  and\ \citenamefont {Schwab}}]{bae_2014}%
  \BibitemOpen
  \bibfield  {author} {\bibinfo {author} {\bibfnamefont {J.}~\bibnamefont
  {Suh}}, \bibinfo {author} {\bibfnamefont {A.~J.}\ \bibnamefont {Weinstein}},
  \bibinfo {author} {\bibfnamefont {C.~U.}\ \bibnamefont {Lei}}, \bibinfo
  {author} {\bibfnamefont {E.~E.}\ \bibnamefont {Wollman}}, \bibinfo {author}
  {\bibfnamefont {S.~K.}\ \bibnamefont {Steinke}}, \bibinfo {author}
  {\bibfnamefont {P.}~\bibnamefont {Meystre}}, \bibinfo {author} {\bibfnamefont
  {A.~A.}\ \bibnamefont {Clerk}},\ and\ \bibinfo {author} {\bibfnamefont
  {K.~C.}\ \bibnamefont {Schwab}},\ }\bibfield  {title} {\bibinfo {title}
  {Mechanically detecting and avoiding the quantum fluctuations of a microwave
  field},\ }\href {https://doi.org/10.1126/science.1253258} {\bibfield
  {journal} {\bibinfo  {journal} {Science}\ }\textbf {\bibinfo {volume}
  {344}},\ \bibinfo {pages} {1262} (\bibinfo {year} {2014})},\ \Eprint
  {https://arxiv.org/abs/https://www.science.org/doi/pdf/10.1126/science.1253258}
  {https://www.science.org/doi/pdf/10.1126/science.1253258} \BibitemShut
  {NoStop}%
\bibitem [{\citenamefont {Shomroni}\ \emph {et~al.}(2019)\citenamefont
  {Shomroni}, \citenamefont {Qiu}, \citenamefont {Malz}, \citenamefont
  {Nunnenkamp},\ and\ \citenamefont {Kippenberg}}]{bae_meas_2019}%
  \BibitemOpen
  \bibfield  {author} {\bibinfo {author} {\bibfnamefont {I.}~\bibnamefont
  {Shomroni}}, \bibinfo {author} {\bibfnamefont {L.}~\bibnamefont {Qiu}},
  \bibinfo {author} {\bibfnamefont {D.}~\bibnamefont {Malz}}, \bibinfo {author}
  {\bibfnamefont {A.}~\bibnamefont {Nunnenkamp}},\ and\ \bibinfo {author}
  {\bibfnamefont {T.~J.}\ \bibnamefont {Kippenberg}},\ }\bibfield  {title}
  {\bibinfo {title} {Optical backaction-evading measurement of a mechanical
  oscillator},\ }\href {https://doi.org/10.1038/s41467-019-10024-3} {\bibfield
  {journal} {\bibinfo  {journal} {Nature Communications}\ }\textbf {\bibinfo
  {volume} {10}},\ \bibinfo {pages} {2086} (\bibinfo {year}
  {2019})}\BibitemShut {NoStop}%
\end{thebibliography}

%

\end{document}